\documentclass[twocolumn, reprint, secnumarabic, superscriptaddress, showkeys,
amssymb, amsmath, nobibnotes, aps,prl]{revtex4-2}

\usepackage{graphicx}
\usepackage{dcolumn}
\usepackage{bm}
\usepackage{physics}

\begin{document}

\title[Hfq--NTR dynamics]{Mobility of bacterial protein Hfq on dsDNA; Role of C-terminus \\ mediated transient binding}

\author{Chuan Jie Tan}
\author{Rajib Basak}
\author{Indresh Yadav}
\author{Jeroen A. van Kan}
\affiliation{Department of Physics, National University of Singapore, Singapore 117542}
\author{V\'eronique Arluison}
\affiliation{Universit\'e de Paris, UFR SDV, 75006 Paris, France}
\affiliation{Laboratoire L\'eon Brillouin, CEA, CNRS, Universit\'e Paris Saclay, 91191 Gif-sur-Yvette, France}
\author{Johan R. C. van der Maarel}
\email{johanmaarel@gmail.com}
\affiliation{Department of Physics, National University of Singapore, Singapore 117542}

\date{\today}

\begin{abstract}
{The mobility of protein is fundamental in the machinery of life. Here, we have investigated the effect of DNA binding in conjunction with DNA internal motion of the bacterial Hfq master regulator devoid of its amyloid C-terminus domain. Hfq is one of the most abundant nucleoid associated proteins that shape the bacterial chromosome and is involved in several aspects of nucleic acid metabolism. Fluorescence microscopy has been used to track a C-terminus domain lacking mutant form of Hfq on double stranded DNA, which is stretched by confinement to a rectangular nanofluidic channel. The mobility of the mutant is strongly accelerated with respect to the wild type variant. Furthermore, it shows a reverse dependence on the internal motion of DNA, in that slower motion results in slower protein diffusion. Results demonstrate the subtle role of DNA internal motion in controlling the mobility of a nucleoid associated protein, and, in particular, the importance of transient binding and moving DNA strands out of the way.}
\begin{center}
\includegraphics[height=3.5 cm]{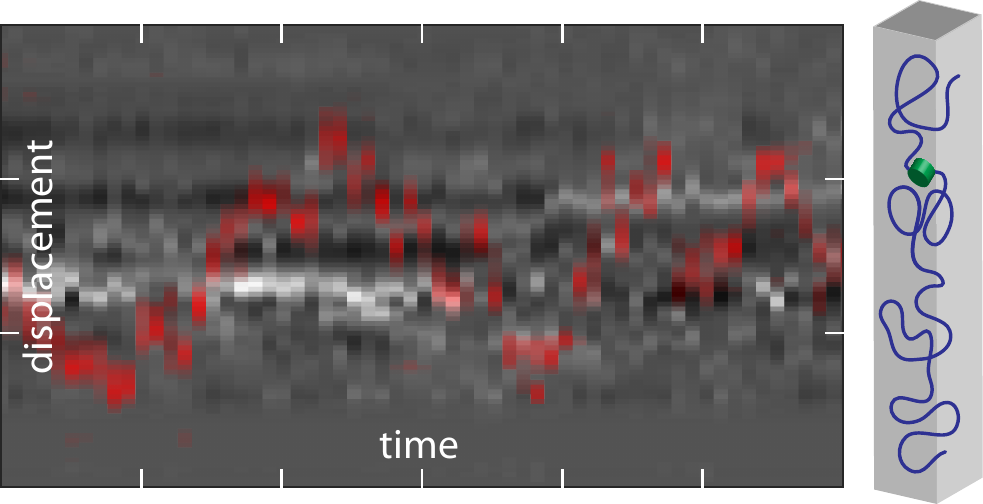}
\end{center}
\end{abstract}


\maketitle

\section*{Introduction}
The mobility of protein impacts DNA metabolism including replication, repair, and gene expression regulation. Accordingly, it has been investigated with a myriad of experimental methodologies and theoretical approaches. In particular, proteins moving on DNA have been visualized and tracked in real time with fluorescence microscopy \cite{kabata1993visualization,Wang2006, blainey2006base, graneli2006long, kim2007single, bonnet2008sliding, davidson2016rapid, vestergaard2018single}. These and other investigations have shown that target search is facilitated by a combination of 1D sliding along DNA and 3D hopping and jumping between different binding sites of DNA \cite{Hu_2007, gorman2008, lomholt2009, hammar2012lac, mahmutovic2015matters}. Another important, yet often ignored aspect of the dance of protein with DNA is the role of internal DNA motion \cite{Mondal, tan2016dynamic, Chow:2017aa}. Previously, we have shown that the mobility of bacterial nucleoid associated protein (NAP) Hfq shows a strong dependence on the internal motion of double stranded DNA, in that slower DNA motion results in faster protein diffusion \cite{Yadav2020uq}. A model of intermittent diffusion was proposed, based on 3D diffusion through the interior of the DNA coil interspersed by periods in which the protein is immobilized in a bound state. More precisely, the protein diffusion coefficient was predicted and confirmed to be proportional to the relaxation (Rouse) time of DNA. The model also predicts that the diffusion coefficient is inversely proportional to the residence time of bound protein on DNA. The present contribution seeks to further explore the role of transient binding by using a mutant form of Hfq devoid of its amyloid C-terminus domain, that is the Sm hexameric torus (residues 1-72, referred to as Hfq--NTR).

Hfq is one of the most abundant NAPs that shape the bacterial chromosome. Its cellular content is comparable to the ones of Fis (factor for inversion simulation) and HU (heat-unstable protein) \cite{AliAzam1999}. Hfq was initially described in relation to its role in bacteriophage Q$\beta$ RNA replication (Hfq stands for host factor for phage Q$\beta$) \cite{fernandez1972}. Later, it was found to play a role in {\it E. Coli} RNA metabolism \cite{vogel2011}. A few studies have been devoted to the role of Hfq in DNA metabolism \cite{lederout2010, cech2014, Cech:2016aa}. In {\it vitro} studies have shown that Hfq forms a nucleoprotein complex and changes the mechanical properties of the double helix \cite{jiang2015,malabirade2017}. Furthermore, protein mediated bridging of DNA segments results in compaction of DNA into a condensed form. It was also found that the C-terminal domain assembles DNA into an elastic gel \cite{El-Hamoui:2020vp}. Hfq is part of a family of NAPs that bridges sections of the DNA molecule, thereby organizing large parts of the genome into chromosomic domains \cite{dorman2009,LIOY2018771}. The propensity for bridging is related to the multiarm functionality of the Hfq hexamer, resulting from self-assembly of the C-terminal domains and binding to the duplex \cite{jiang2015}. Indeed, for Hfq--NTR bridging and compaction of DNA into a condensed form was not observed \cite{malabirade2017}. Nevertheless, the torus still binds on DNA but with an order of magnitude shorter residence time as compared to full length Hfq. It is thus expected that the lower binding affinity has a profound effect on protein diffusivity.
 
In the present contribution, fluorescence microscopy is used to track Cy3-labeled Hfq--NTR on dsDNA. For a description of the methodology, our previous report on the transport of full length Hfq on DNA in otherwise the same experimental conditions may be consulted \cite{Yadav2020uq}. The untethered DNA molecules are stretched by confinement to a long, rectangular channel with a diameter of 125 nm. Single proteins will be hosted by bacteriophage $\lambda$-DNA and its di- and trimeric concatemers. In the same channel system, Rouse relaxation times pertaining to internal DNA motion were previously measured through analysis of fluorescence correlation \cite{yadav2020}. For the longer concatemers, internal motion is considerably slowed down due to the cubic dependence of the Rouse time on the molecular weight. The role of transient binding will then be investigated by comparing the diffusion coefficients of full length Hfq and its Hfq--NTR mutant and their dependencies on the Rouse time of the hosting DNA molecule.

\begin{table*}
\begin{center}
\caption{Diffusion coefficient $D$ of Hfq and Hfq--NTR inside a 125-nm channel without DNA as well as diffusing on $\lambda$-DNA and its di- and trimeric concatemers. DNA Rouse time $\tau_R$, stretch $R$, dye-corrected contour length $L$, and stretch per unit contour length $R/L$ are also included \cite{yadav2020, Yadav2020uq}.}
{\footnotesize{
\begin{tabular}{lll l l l l}
\hline\noalign{\smallskip}
& $D$ ($\mu$m$^2$/s) & $D$ ($\mu$m$^2$/s) & $\tau_R$ (ms) & $R$ ($\mu$m) &$L$ ($\mu$m)& $R/L$\\
&Hfq&Hfq--NTR& & && \\
\noalign{\smallskip}
\hline\noalign{\smallskip}
DNA-free& 1.00$\pm$0.02 & 4.2$\pm$0.1&-&-&-&-\\
$\lambda$-DNA & 0.0185$\pm$0.0005 & 2.6$\pm$0.1 & 60$\pm$5 & 11.3$\pm$0.1 &21.4& 0.528$\pm$0.005\\
$\lambda$-DNA$_2$ & 0.095$\pm$0.002 & 2.23$\pm$0.08 & 560$\pm$60 & 22.7$\pm$0.2 & 42.8 &0.530$\pm$0.005\\
$\lambda$-DNA$_3$ & 0.30$\pm$0.01 & 1.75$\pm$0.07& 1800$\pm$300 & 29.9$\pm$0.6 &64.2 &0.466$\pm$0.009\\
\noalign{\smallskip}
\hline
\end{tabular}}}
\label{rouse:table}
\end{center}
\end{table*}

\section*{Materials and Methods}
\subsection*{Chip fabrication} 
Nanofluidic chips with channels of length 90 ${\mu}$m, depth $130 \pm 5$ nm, and width $120 \pm 5$ nm (125-nm channel system) were fabricated by replication in polydimethylsiloxane with enhanced elasticity modulus (X-PDMS) of a patterned master stamp \cite{vankan2006,zhang2008, vankan2012}. The nanochannel part of the stamp was made in hydrogen silsesquioxane resist (Dow Corning, Midland, MI) using a lithography process with proton beam writing. An array of nanochannels is connected to two loading reservoirs through a superposing set of microchannels made in mr-DWL photoresist (Micro Resist Technology, Berlin, Germany) with a aser writer (Heidelberg micro PG 101). The heights and widths of the ridges in the master stamp were measured with atomic force microscopy (Dimension 3000, Veeco, Woodbury, NY) and scanning electron microscopy (JEOL JSM6700F), respectively. The master stamp was coated with diamond-like-carbon layer before being copied in the inorganic-organic hybrid polymer OrmoStamp (Micro Resist Technology). After copying, the stamp was coated with a 5-nm-thick carbon layer to ensure release of the replicated chips. The stamp was replicated in X-PDMS followed by curing at 333 K for 24h. Following plasma oxidation (Harrick, Ossining, NY), the X-PDMS replica was sealed with a glass coverslip. 
		
\subsection*{Sample preparation} 
$\lambda$-DNA (48.5 kbp, contour length of 16.5 $\mu$m) was purchased from New England Biolabs, Ipswich, MA. The stock solution has a concentration of 0.5 g of DNA/L in TE buffer (10 mM Tris-HCl, pH 8.0, and 1 mM EDTA). Covalently bonded concatemers ($\lambda$-DNA$_2$ dimers and $\lambda$-DNA$_3$ trimers) of $\lambda$-DNA were prepared by joining the cohesive ends through phosphodiester bonds and enzymatic ligation with T4 DNA ligase (Promega, Madison, WI) \cite{bauer2017}. The appropriate amounts of DNA and enzyme were mixed and ligation was carried out overnight at 277 K. The reaction was inactivated by heating the sample to 338 K for 10 min. Finally, the sample was dialyzed in a micro-dialyzer to remove excess salts and enzyme before re-dispersing in TE buffer. The peptide corresponding to the Hfq-NTR domain (residues 1-72) was purchased from Proteogenix SA (France). To obtain the fluorescent form of the protein, a S38C mutation was introduced and labeled with a Cy3 maleimide reactive-dye (GE-Healthcare, Chicago, IL) \cite{rabhi2011sm}. The labeling efficiency is 80\%, that is the average number of fluorphores per Hfq--NTR hexamer is about five. Folding of the peptide was confirmed with UV circular dichroism spectroscopy. Solutions of DNA and Hfq--NTR were mixed and incubated overnight at 277 K. The final concentrations are 0.3 mg of DNA/L and 0.01 mg of Hfq--NTR/L. Prior to fluorescence imaging, DNA was stained with intercalating dye YOYO-1 (Invitrogen, Carlsbad, CA, USA) at a ratio of one dye molecule for ten base pairs. No anti-photo bleaching agent was used. 

\subsection*{Fluorescence imaging} 
The stained DNA/Hfq--NTR solution was pipeted into the loading reservoirs connected by the array of nanochannels. The DNA molecules were subsequently driven into the channels by electrophoresis. For this purpose, two platinum electrodes were immersed in the reservoirs and connected to a power supply with a voltage in the range 0.1--10 V (Keithley, Cleveland, OH). Once the DNA molecules were brought inside the nanochannels, the electric field was switched off and the molecules were allowed to relax to their equilibrium state for 2--5 min. A fresh chip was used for measurement of about 5 molecules and the experiments were done at an ambient temperature of 296 K. Protein and DNA were visualized with a Nikon Eclipse Ti inverted fluorescence microscope equipped with 200-mW/488-nm (YOYO-1), 400 mW/556-nm (Cy3), and 200 mW/640-nm lasers, a 405/488/561/640-nm filter cube (Chroma Technology, Bellows Falls, VT), and 100$\times$ oil immersion objective (numerical aperture 1.49). Imaging of DNA and Hfq--NTR was facilitated by switching and selecting the relevant laser excitation wavelength for YOYO-1 and Cy3, respectively. Video was recorded with an electron-multiplying charged coupled device camera (Andor iXon X3). The image pixel size of $0.16 \times 0.16$ $\mu$m$^2$ was calibrated with the help of a metric ruler. Hfq--NTR trajectories, mean square displacements, and diffusion coefficients were obtained with home-developed scripts in {\sc matlab}, R2021b (MathWorks, Natick, MA).

\begin{figure}
\begin{center}
\includegraphics[width=8cm]{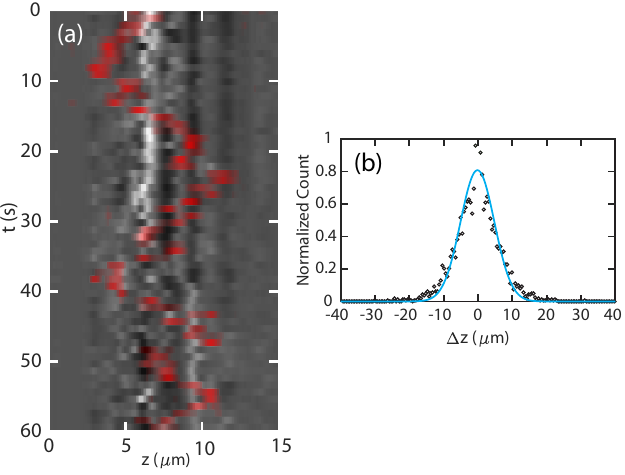}
\caption{(a) Kymograph of $\lambda$-DNA density fluctuation (gray scale). The simultaneously recorded trajectory of a Hfq--NTR protein diffusing on the DNA molecule is superposed in red. For each dye the exposure time is 500 ms, resulting in a rate of 1 frame per s. The duration of the clip is 1 min. (b) Probability distribution for the displacement $\Delta z$ ($\circ$) along the direction of the channel of Hfq--NTR on $\lambda$-DNA$_2$ for lag-time $\tau$ = 6 s. The solid curve represents a Gaussian fit.}
\label{ntrdyn:fig1} 
\end{center}
\end{figure}

\begin{figure}
\begin{center}
\includegraphics[width=8.5 cm]{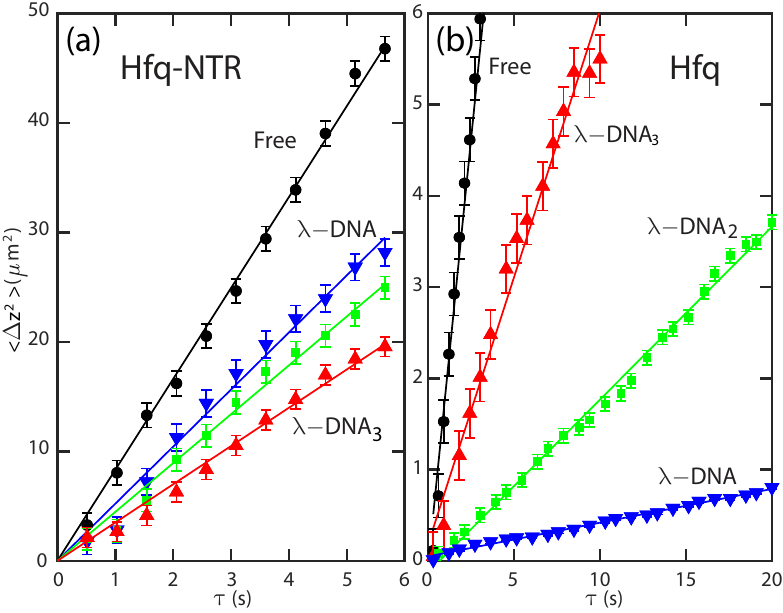}
\caption{(a) Mean square displacement $\left < \Delta z^2 \right >$ for free Hfq--NTR ($\circ$, black), Hfq--NTR on $\lambda$-DNA ($\triangledown$, blue), $\lambda$-DNA$_2$ ($\square$, green), and $\lambda$-DNA$_3$ ($\triangle$, red) inside the channel (average over 13, 19, 21, and 23 molecules, respectively). The solid lines represent linear fits giving the diffusion coefficient $D$. (b) As in panel (a) but for Hfq (average over 15, 20, 20, and 15 molecules, respectively). Data for Hfq are from Ref. \citenum{Yadav2020uq}.}
\label{ntrdyn:fig2} 
\end{center}
\end{figure}

\section{Results and Discussion}
DNA and Hfq--NTR were imaged by recording video clips with frame-alternating excitation of DNA-bound YOYO-1 and protein-bound Cy3 dye, with an exposure time of 500 ms, and a rate of 1 frame per second per dye. For representative fluorescence images of DNA in the same 125 nm channel system, Refs. \citenum{yadav2020} and \citenum{Yadav2020uq} may be consulted. $\lambda$-DNA and its dimeric $\lambda$-DNA$_2$ and trimeric $\lambda$-DNA$_3$ concatemers can easily be discerned by their stretch. Results for the stretch as well as the dye-corrected lengths along the contour of the DNA molecules are collected in Table \ref{rouse:table}. With a stretch of about half the contour length, the molecules are coiled but close to the deflection regime as indicated by Monte Carlo simulation \cite{Dai2016}. A coiled conformation is in accordance with the 125 nm diameter of the channel, which is about two times the stained DNA persistence length of 60 nm \cite{Kundukad2014}. Furthermore, the stretch per unit length of the contour $R/L$ and, hence, DNA segment density are almost constant. For the trimeric $\lambda$-DNA$_3$ concatemer, the value of $R/L$ is less but within 12\% of the value of the monomers. The correspondingly higher DNA density is of no consequence for the analysis of the protein diffusion data.

The fluctuation in density of a single $\lambda$-DNA molecule as measured through YOYO-1 fluorescence projected along the channel is displayed in Fig. \ref{ntrdyn:fig1}(a). Such a kymograph serves to characterize internal DNA motion, as has previously been described \cite{yadav2020}. Superposed is the simultaneously measured trajectory of a Hfq--NTR protein, which is diffusing inside the channel in the same space occupied by the hosting DNA molecule. The protein moves along the DNA molecule over several micrometers but does not disengage during the time the video is recorded. Significant sticking and immobilization of the protein on the surface of the nanofluidic device was not observed. Once the presence of a single protein on DNA was established, its trajectory was recorded with exclusive excitation of the Cy3 dye at a rate of 2 frames per second and clip duration of 2--5 min.

\begin{figure}
\begin{center}
\includegraphics[width=8.5cm]{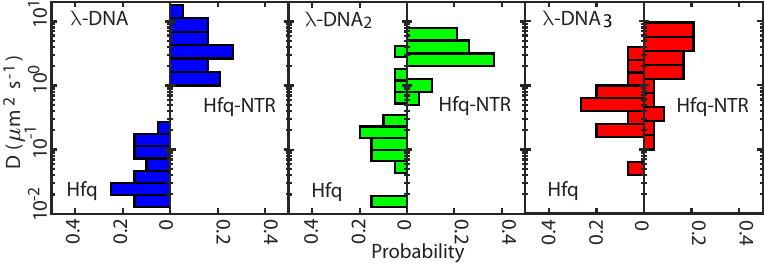}
\caption{Histogram violin plots of diffusion coefficients $D$ pertaining to Hfq and Hfq--NTR on $\lambda$-DNA (blue), $\lambda$-DNA$_2$ (green), and $\lambda$-DNA$_3$ (red) from left to right.}
\label{ntrdyn:fig3} 
\end{center}
\end{figure}

The trajectories of a total of 19, 21, and 23 proteins diffusing on $\lambda$-DNA, $\lambda$-DNA$_2$, and $\lambda$-DNA$_3$, respectively, were analyzed. From the pooled trajectories pertaining to the same DNA host, the probability distributions for displacement of Hfq--NTR in the longitudinal direction of the channel were determined for a range of lag-times $\tau$. The minimum lag-time of 500 ms is determined by the exposure time required for imaging of the Cy3 dye. The practical limit of the maximum lag-time is 6 s due to the relatively high mobility and finite duration of the video clips. An example of the probability distribution for the displacement is shown in Fig. \ref{ntrdyn:fig1}(b). For each lag-time, a Gaussian was fitted to the probability distribution by optimizing the mean square displacement $\left < \Delta z^2 (\tau) \right >$. The results for Hfq--NTR are shown in Fig. \ref{ntrdyn:fig2}(a). For comparison, the previously reported results for full length Hfq are displayed in Fig. \ref{ntrdyn:fig2}(b). Note that $\left < \Delta z^2 \right >$ refers to one-dimensional diffusion of the protein in the longitudinal direction of the channel and in the interior of the indicated DNA host. Linear least-squares fits intersecting the origin confirm diffusional transport according to $\left < \Delta z^2 \right > = 2 D\tau$ with diffusion coefficient $D$. Values of $D$ and their standard deviations for proteins diffusing on the various DNA hosts as well as for proteins diffusing inside the channel without DNA are collected in Table \ref{rouse:table}. Striking observations are 1--2 orders of magnitude higher values of $D$ and the less pronounced, reverse dependence on host molecular weight for Hfq--NTR. These results clearly demonstrate the pivotal role of the C-terminal domain in determining Hfq's mobility on DNA.

Diffusion coefficients of individual proteins can be obtained through analysis of single trajectories. Histogram violin plots showing the results for Hfq and Hfq--NTR are displayed in Fig. \ref{ntrdyn:fig3}. Irrespective of the DNA host, variations over 1--2 orders of magnitude in values of $D$ are observed. These variations are similar to those reported for other DNA binding proteins \cite{Wang2006, blainey2006base, graneli2006long, kim2007single, bonnet2008sliding, davidson2016rapid, vestergaard2018single}. The observed trend towards higher values of the diffusion coefficient and the less pronounced, reverse dependence on DNA molecular weight for Hfq--NTR are in agreement with the ensemble pooled results. Variation in mobility of individual DNA binding proteins can be reconciled by distributions in DNA Rouse time and, base pair sequence-dependent, protein residence time.

\begin{figure}[t]
\begin{center}
\includegraphics[width=5 cm]{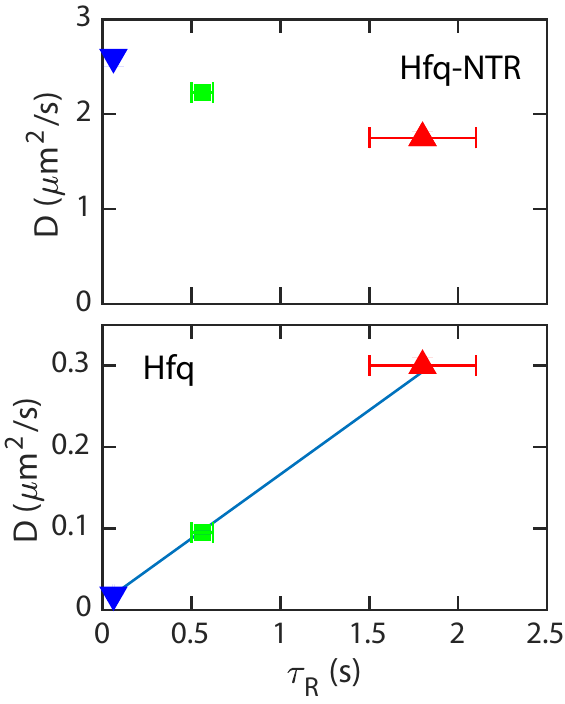}
\caption{Hfq--NTR (top) and Hfq (bottom) diffusion coefficient $D$ as derived from the pooled trajectories vs DNA Rouse time $\tau_R$ hosted by $\lambda$-DNA (blue), $\lambda$-DNA$_2$ (green), and $\lambda$-DNA$_3$ (red). The solid line for Hfq represents $D \propto \tau_R$ variation.}
\label{ntrdyn:fig4} 
\end{center}
\end{figure}

Internal motion of $\lambda$-DNA and its concatemers in exactly the same channel system was previously investigated through analysis of fluorescence correlation \cite{yadav2020}. The results agree with Rouse dynamics due to screening of hydrodynamic interaction beyond a distance of around the diameter of the channel (125 nm). The reported values of the relaxation time associated with fluctuation in DNA density, that is Rouse time $\tau_R$, are also collected in Table \ref{rouse:table}. Following a threefold increase in molecular weight, intramolecular dynamics is slowed down considerably with a two orders of magnitude increase in value of $\tau_R$. There is however no substantial change in time-averaged density because the stretch remains proportional to the contour length within a margin of 12\%. In the case of full length Hfq, a linear relation between the diffusion constant and Rouse time of the hosting DNA molecule was previously reported \cite{Yadav2020uq}. For Hfq--NTR, the situation is fundamentally different. Besides the fact that the values of the diffusion coefficients are 1--2 orders of magnitude higher than the ones pertaining to Hfq, they also decrease with increasing Rouse time (slower internal motion) of the DNA host (see Fig. \ref{ntrdyn:fig4}).

Protein transport on DNA is usually envisioned by a combination of 1D sliding along DNA and 3D diffusion between different binding sites of DNA \cite{Hu_2007, gorman2008, lomholt2009, hammar2012lac, mahmutovic2015matters}. The DNA molecules of various molecular weight inside the channel are not stretched to full extent but are coiled with an almost fixed stretch over contour length ratio of about 0.5 (see Table \ref{rouse:table}). In the case of 1D sliding along DNA, the diffusion coefficient should, hence, not depend on the molecular weight of the hosting molecule. For full length Hfq as well as its mutant, there is a significant dependence of the diffusion coefficient on DNA molecular weight. Accordingly, the 1D sliding mechanism is of minor importance for interpretation of the present data. The typical displacements of the protein cover the stretch of the DNA molecule, whereas the DNA segments are confined to a region with a size of around the channel diameter (125 nm). The protein cannot be bound to DNA for longer periods of time, because DNA moves very slowly and sub-diffusion is not observed (see Fig. \ref{ntrdyn:fig2}). The most probable mechanism is 3D diffusion through the interior of the DNA coil, possibly interspersed by periods in which the protein is immobilized in a bound state.

In previous work, we proposed a model of intermittent diffusion, that is DNA binds the protein and the protein can only make an excursion through the interior of the coil if it disengages from its binding site \cite{Yadav2020uq}. A key element is that the duration of the excursion is set by the timescale of DNA density fluctuation, that is Rouse time $\tau_R$. The diffusion coefficient then takes the form $ D \sim D_f \tau_R /\tau_B$, with $D_f$ being the diffusion coefficient in the interior of the coil and $\tau_B$ the residence time of bound protein on DNA. For full length Hfq, a linear dependence of $D$ on $\tau_R$ is indeed observed (see Fig. \ref{ntrdyn:fig4}, bottom). The precise value of $D_f$ is unknown but is expected to be close to the value measured for the longest concatemer, that is $D_f \sim 0.3$ $\mu$m$^2$/s. With this value of $D_f$, the residence time of bound full length Hfq on DNA, $\tau_B \sim$ 2 s. A bound residence time of a few seconds agrees with residence times ranging between seconds and minutes of many transcription factors on DNA \cite{azpeitia_short_2020}. 

Values of the diffusion coefficient pertaining to the mutant exceed the ones for full length Hfq by 1--2 orders of magnitude depending on DNA molecular weight. Furthermore, the diffusion coefficient decreases rather than increases with increasing Rouse time (see Fig. \ref{ntrdyn:fig4}, top). In particular, the latter observation is at odds with the $D \propto \tau_R$ prediction of the intermittent diffusion model. Protein molecular weight cannot be a conclusive factor, considering the moderate difference in mobility in the absence of DNA. A plausible explanation for accelerated diffusion is the lack of bridging interaction and the order of magnitude lower binding affinity of the mutant \cite{malabirade2017}. Without binding on DNA for longer times, protein transport may be described by the gated diffusion model \cite{cai_hopping_2015,Chow:2017aa}. In the case of gated diffusion, DNA confines the protein in a cage and the protein can move between cages when hindering DNA strands move out of the way. The gated diffusion model predicts accelerated diffusion with faster internal DNA motion (shorter Rouse time). Gated diffusion does not agree with our observations for full length Hfq. However, in the case of less prominent binding of the mutant, the gated diffusion model comes to the fore and correctly predicts a decreasing value of the diffusion coefficient with increasing Rouse time. 

\section{Conclusions}
With fluorescence microscopy, we have tracked a C-terminus domain-lacking mutant of Hfq on DNAs of various molecular weight. The DNA molecules are stretched inside a channel with a diameter of about two times the persistence length. There is no significant variation in DNA density, because the stretch is almost proportional to the contour length. There is, however, an almost two orders of magnitude variation in relaxation time pertaining to internal DNA motion, because of its cubic dependence on molecular weight \cite{yadav2020}. Values of the diffusion coefficient pertaining to individual proteins show a wide variation, as is the case for full length Hfq and many other DNA binding proteins \cite{Wang2006, blainey2006base, graneli2006long, kim2007single, bonnet2008sliding, davidson2016rapid, vestergaard2018single}. However, as compared to previously reported results for full length Hfq, the ensemble pooled values of the diffusion coefficients for the mutant are 1--2 orders of magnitude higher and they decrease rather than increase with slower internal DNA motion. These observations unambigeously demonstrate the pivotal role of C-terminal mediated DNA binding, because the mutant has a lower binding affinity and lacks the propensity for bridging distal DNA segments \cite{malabirade2017}. 

Previously, the linear dependence of full length Hfq's diffusion coefficient on DNA's relaxation time was explained with a intermittent diffusion model based on 3D diffusion through the interior of the DNA coil interspersed by periods, in which the protein is immobilized in a bound state \cite{Yadav2020uq}. For the mutant without binding on DNA for longer times, the diffusion can be described by the gated diffusion model \cite{cai_hopping_2015,Chow:2017aa}. The latter model predicts accelerated diffusion with faster internal DNA motion, because hindering DNA strands need to move out of the way. The present study demonstrates the subtle role of internal DNA motion in controlling the mobility of a nucleoid associated protein, and, in particular, the importance of transient binding. Furthermore, it highlights the role of Hfq's amyloid region interaction with DNA, and in general the role of amyloids in DNA-related processes \cite{Cordeiro:2014vk}. For instance, they could play a role in the toxicity of human amyloids responsible for neurological disorders, such as Alzheimer's A$\beta$ disease. Indeed, these amyloids bind to DNA, have been found inside the nucleus, and may strongly affect genetic expression \cite{jimenez2010protein, barucker2014nuclear}.

\section{ACKNOWLEDGEMENTS}
This work was supported by Ministry of Education, Singapore (MOE) Academic Research Fund Tier 1 Grants R144000414114 and R144000451114.
\vspace{-0.2 cm}
%
%
%


\providecommand{\noopsort}[1]{}\providecommand{\singleletter}[1]{#1}%

\end{document}